\pdfoutput=1
\documentclass[%
reprint,
superscriptaddress,
showpacs,preprintnumbers,
showkeys,
nofootinbib,
amsmath,amssymb,
aps,
pra,
]{revtex4-1}

\usepackage{comment}
\usepackage{bm}% bold math
\usepackage{gensymb}
\usepackage{graphicx}% Include figure files
\usepackage{dcolumn}% Align table columns on decimal point
\usepackage{gensymb}
\usepackage{hyperref}

\usepackage{enumitem}
\hypersetup{%
	colorlinks = true,
	citecolor = blue,
	linkcolor = blue,
	urlcolor = blue,
	linkbordercolor = 0 0 0,
	pdfborder= 0 0 0,
}

\usepackage[pagewise, mathlines]{lineno}% Enable numbering of text and display math
\usepackage{multirow}
\usepackage{physics}

\usepackage{svg}
\usepackage{tikz}
\usepackage[normalem]{ulem}
\usepackage{xcolor}

\newcommand{\p}{\partial}

\newcommand{\Calcium}{{}^{40}\mathrm{Ca}^+}
\newcommand{\Strontium}{{}^{88}\mathrm{Sr}^+}
\newcommand{\Barium}{{}^{138}\mathrm{Ba}^+}
\newcommand{\Radium}{{}^{226}\mathrm{Ra}^+}

\DeclareMathOperator{\Arg}{Arg}

\begin{document}
	
	\preprint{}
	
  \title{Individual addressing of ion qubits with counter-propagating optical frequency combs}
	
  \author{Evgeny Anikin}
  \author{Lianna A. Akopyan}
  \author{Mikhail Popov}
  \author{Yelnury Suleimen}
	\author{Olga Lakhmanskaya}
  \author{Kirill Lakhmanskiy}
	\affiliation{Russian Quantum Center,  Skolkovo, Moscow 143025, Russia}

	\date{\today}% It is always \today, today,
	
\begin{abstract}
  
%  Individual addressing is an important challenge for scalability of the trapped-ion quantum computers. 
We propose a new method of individual single-qubit addressing of linear trapped-ion chains utilizing two ultrastable femtosecond frequency combs. For that, we suggest implementing the single-qubit gates with two counter-propagating frequency combs overlapping on the target ion and causing the AC Stark shift between the qubit levels. With analytical calculations and numerical modeling, we show that the arbitrary single-qubit rotations can be indeed realized using only laser fields propagating along the ion chain. We analyze the  error sources for the proposed addressing method and prove that it allows implementing the single-qubit gates with high fidelity.

%We propose a new method of individual addressing of trapped ions using two counter-propagating frequency combs with pulses meeting on the addressed ion. For this we derive theoretically and model numerically an ac-Stark shift induced by the combs within and outside of the overlap region of the combs. 
%We show that arbitrary one-qubit rotations can be indeed realised using only global addressing. We analyse possible error sources for the proposed gate and estimate its fidelity.
%%Exact setup parameters for $^{40}Ca$ trapped ion qubits are presented with characteristic gate times of $2$ $\mu$s and the infidelity sources being at the level of $10^{-4}$. 
%

\end{abstract}

\maketitle

\section{Introduction}

Trapped-ion quantum computers are one of the most promising platforms for quantum computation \cite{Bermudez2017, Bruzewicz2019}. 
%\cite{Bermudez2017}%, Bruzewicz2019}. 
Their beneficial features include long qubit coherence times \cite{Haeffner2005, Wang2021}, 
high entangling gate fidelities \cite{Clark2021}, and all-to-all qubit connectivity.
With trapped-ion quantum computers, highly entangled states have been prepared \cite{Pogorelov2021}, 
and quantum circuits consisting of multiple gates have been realized \cite{Debnath2016, Wright2019}.
Also, trapped ions allowed performing quantum simulations of various spin models with up to 53 spins \cite{Richerme2013a, Smith2016, Zhang2017, Monroe2021}.
However, scaling the trapped ion quantum computers up to more than tens of qubits remains challenging \cite{Monroe2013}.
%difficulties including higher heating rate and imperfect control of motional state.

One of the necessary components to perform quantum operations with trapped ions
is individual addressing, meaning the ability to apply the control laser 
field to an individual ion to perform gate operations. The existing approaches to perform addressing include the usage of micro-optics splitting modules 
\cite{Day2021,Pogorelov2021}, acousto-optical deflectors \cite{Pogorelov2021}, multi-channel acousto-optical modulators \cite{Wright2019}, 
microelectromechanical mirror systems \cite{Wang2020}, or integrated-optical waveguides \cite{Mehta2016}.
%However, these approaches meets challenges when the number of ions in the chain increases. 
However, the difficulty of the technical realization of these approaches increases with the growing number of ions.
Thus, designing new approaches to individual addressing is of great interest to the development of the large scale quantum computer based on trapped ions.
In this manuscript, we suggest a new scalable method of single-qubit addressing in 
trapped-ion quantum computers utilizing the femtosecond frequency combs.

Ultrastable femtosecond frequency combs generated by mode-locked lasers \cite{Cundiff2003} have multiple applications 
in the field of quantum information processing, 
in particular, for quantum computing with atomic ions. For example, the remarkable spectral purity of the frequency combs enabled their usage to produce 
entanglement between two atomic ions via the Raman process \cite{Hayes2010}. Also, the high instantaneous field intensity and the short pulse duration allowed 
the implementation of ultrafast gates for the ions \cite{Campbell2010} and the generation of ultrafast spin-motion entanglement \cite{Mizrahi2013}.

%In most of the trapped ions quantum computers, the ions are arranged in a linear chain \cite{Bruzewicz2019}}.
Most trapped-ion quantum computers are designed implying the arrangement of ions into a linear chain \cite{Bruzewicz2019}.
We suggest a method to perform individual single-qubit operations in linear chains of optical ion qubits with two 
ultrastable femtosecond frequency combs with the same repetition rate 
counter-propagating along the chain. To select the target ion, the delay between the comb pulses should be adjusted to overlap them on the target ion. After the action of the two trains of pulses, the rotation of the target qubit differs from the rotation of the other ones due to the nonlinearity of the ion-field interaction. The beneficial feature of the suggested method is that the setup geometry is independent of the number of ions in the chain.

We demonstrate the feasibility of such an approach for optical $\Calcium$ qubits. We consider the interaction of the $\Calcium$ qubit with the off-resonant comb field inducing the AC Stark shift \cite{Autler1955} on the qubit levels, thus making it possible to implement the $R_z(\theta)$ gate. Our calculations show that the $R_z(\theta)$ gate can be implemented with reasonable gate time and high fidelity. 
Thus, the suggested method can become an effective tool for quantum operations in the ion chain.

\section{Arbitrary local gate based on AC Stark shift}
\label{sec:idea}

\begin{figure}
  \centering
  \begin{minipage}{\linewidth}
    \includegraphics[width=\linewidth]{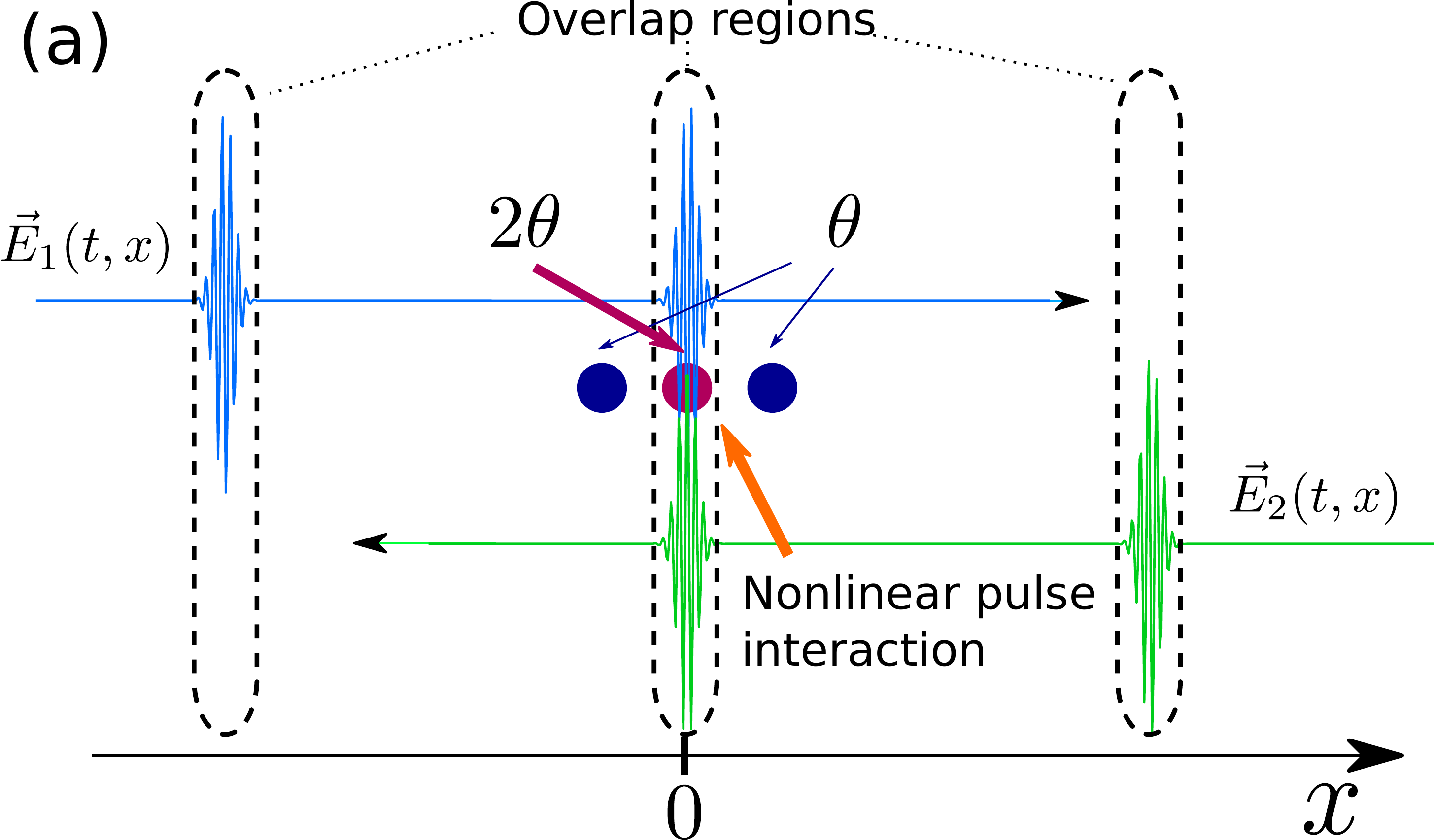} 
  \end{minipage}
  \begin{minipage}{\linewidth}
    \vspace{0.5cm}
    \includegraphics[width=\linewidth]{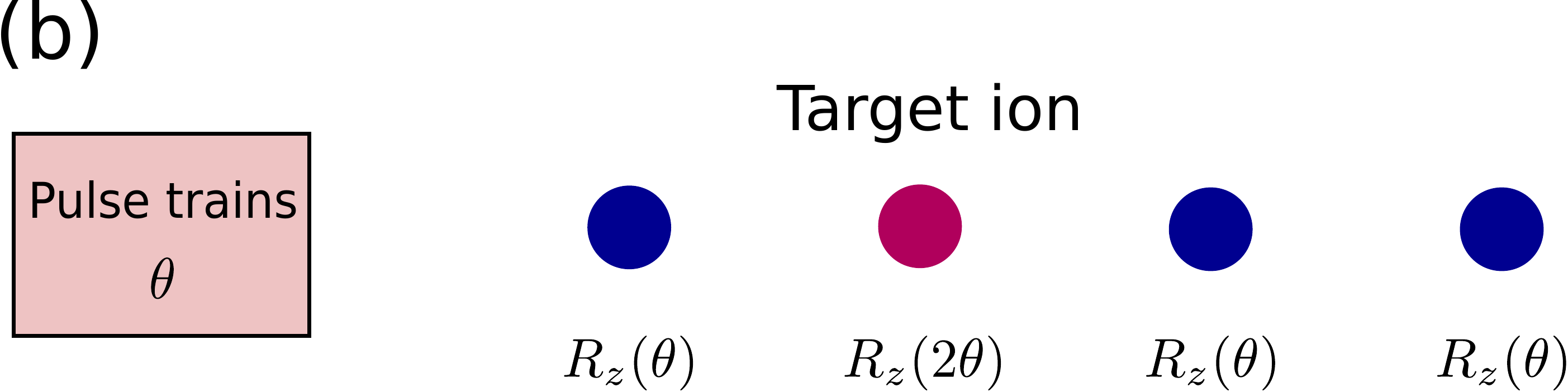}
    \vspace{2mm}
  \end{minipage}
  \begin{minipage}{\linewidth}
    \centering
    \includegraphics[width=\linewidth]{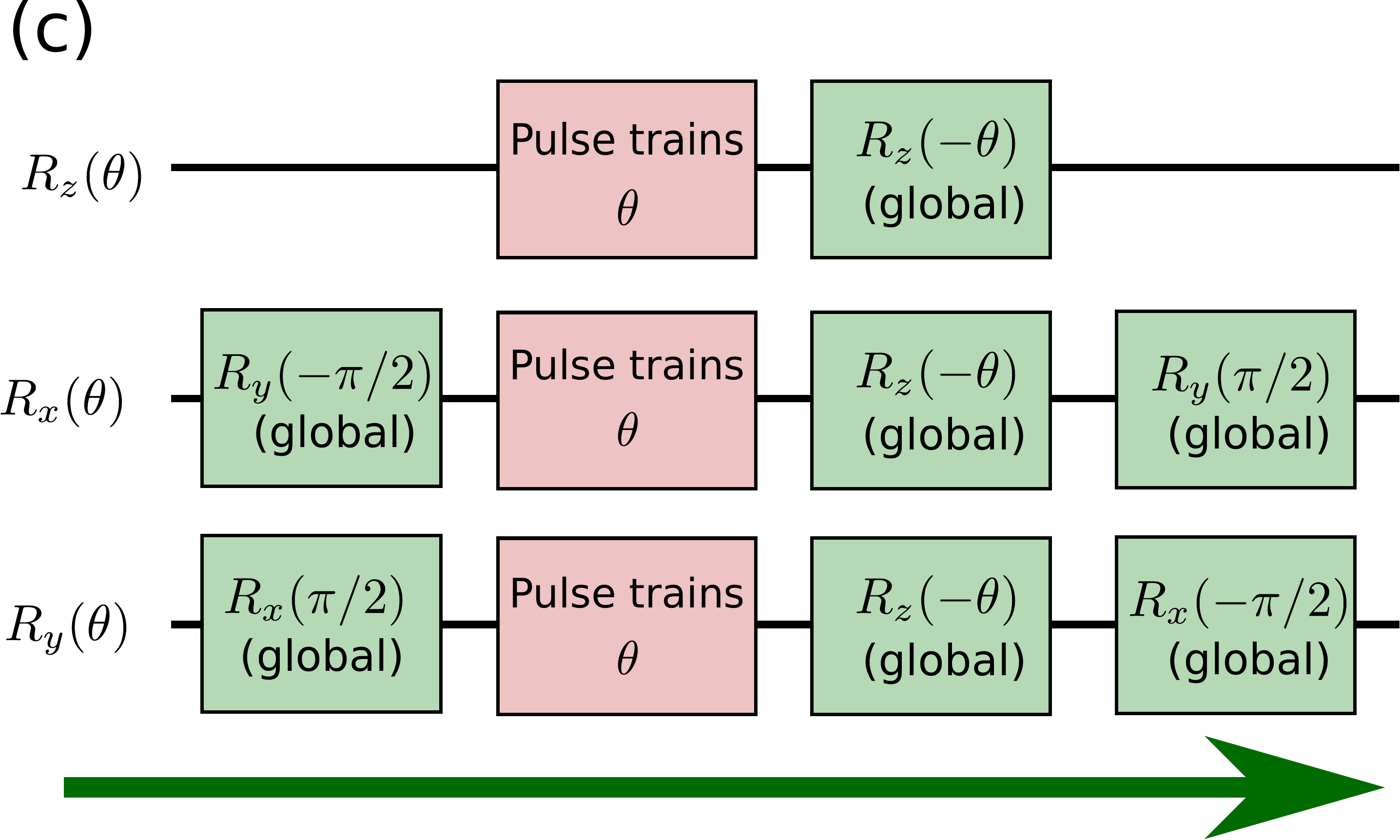}
%    \includesvg[width=0.7\linewidth]{protocols_1.svg}
  \end{minipage}
  \caption{
  (a) Ion chain in the linear trap and comb pulses propagating along it
  ($x$-axis).
  Comb pulses overlap on the target ion.
  (b) The phases acquired by the ion chain after the action of two counter-propagating pulse trains.
  (c) 
  The implementation of the local $R_x(\theta)$, $R_y(\theta)$, $R_z(\theta)$ operations on a target ion.
  The arrow represents the order in which the operators should be performed (from left to right).  
  }
  \label{fig:setup}
\end{figure}

The key idea of our method is to illuminate the ion chain with two frequency combs which propagate along the ion chain in opposite directions 
(see Fig.~\ref{fig:setup}(a)). 
For such geometry, there exist spatial regions  
%(of approximately half the pulses length) 
where the pulses from different combs overlap. 
The length of the femtosecond pulse is of the order of the interionic distance, so it is possible to make only one target qubit lie in the overlap region.
The action of the field on the ion in the overlap region differs from the one outside of it due to the intrinsic nonlinearity in the ion-field interaction.
% For all other ions, the temporal delay between the coming pulses is larger than the pulses duration. 
This allows acting selectively on an arbitrary ion by controlling the time delay between the combs: the time delay should be 
chosen to overlap the pulses on a target ion.

The nonlinear effect that we utilize is the quadratic AC Stark shift of the qubit levels, which
implies a large detuning of the comb frequencies from the ion transition frequencies. Then,
%When the frequency combs are 
%strongly detuned from the ions' transition frequencies, 
the main effect of each comb pulse is the phase accumulation on each of 
the ion levels, in particular, on qubit levels. Therefore, %$k$, 
the action of the comb reduces to the $R_z(\theta)$ gate for each qubit.

At the position of the target ion,
the pulses form an interference pattern. The electric field in the largest interference peak is twice the maximum of the field magnitude of a single pulse. The electric field of the combs induces the phase shifts between the ion qubit levels due to the AC Stark effect. As the AC Stark effect is quadratic in the field amplitude, the phase shift is four times larger than the phase shift from a single pulse. 

Therefore, the cumulative effect of the two trains of $N_\mathrm{pulses}$ 
pulses on the ion qubit levels $|0\rangle$ and $|1\rangle$ 
is an $R_z(\theta)$ rotation. 
The rotation angle is 
$2\theta = 4N_\mathrm{pulses}(\delta\theta_1 - \delta\theta_0)$ 
for the target ion and  $\theta = 2N_\mathrm{pulses}(\delta\theta_1 - \delta\theta_0)$ 
for non-target ions (see Fig.~\ref{fig:setup}(b)), where $\delta\theta_{0,1}$ are the acquired phases per 
single comb pulse for the levels $0$ and $1$, respectively.
To apply the $R_z(\theta)$ rotation only on the target ion, the pulse trains need to be followed by a 
global $R_z(-\theta)$ rotation. Further, local $R_x(\theta)$ and $R_y(\theta)$ rotations can be implemented with the help of additional global $\pm\pi/2$ 
rotations before and after the pulse train (see Fig.~\ref{fig:setup}(c)) \cite{NielsenChuang2010}.

The choice of the comb wavelength is justified by two requirements. First, the combs should be detuned far enough from the transitions between the qubit 
levels and the short-living $4P_{1/2}$ and $4P_{3/2}$ levels, otherwise, the photon scattering will lead to large qubit decoherence. As the spectral width 
of the femtosecond pulses is tenth of terahertz, 
the detunings should be at least the same order or larger.
Second, the comb wavelength should be far enough from the 
magic wavelength of qubit transitions \cite{Jiang2019},  
which ensures that the phase acquired by the qubit is sufficiently large.
For the $\Calcium$, both of these requirements are satisfied for the 1000 nm frequency comb.

The theoretical analysis of the subsequent sections shows that for the $\Calcium$ ion, the
$\sim 1000\,\mathrm{nm}$ frequency combs with a pulse duration of $20\,\mathrm{fs}$ and a repetition rate of $100\,\mathrm{MHz}$ allow implementing
the single-qubit $R_z(\pi/2)$ rotations with the infidelity of $\sim 4\cdot 10^{-4}$ and the gate duration of $\sim 8 \,\mu s$. 

\begin{figure}[h]
  \begin{minipage}{\linewidth}
    \centering
    \includegraphics[width=0.9\linewidth]{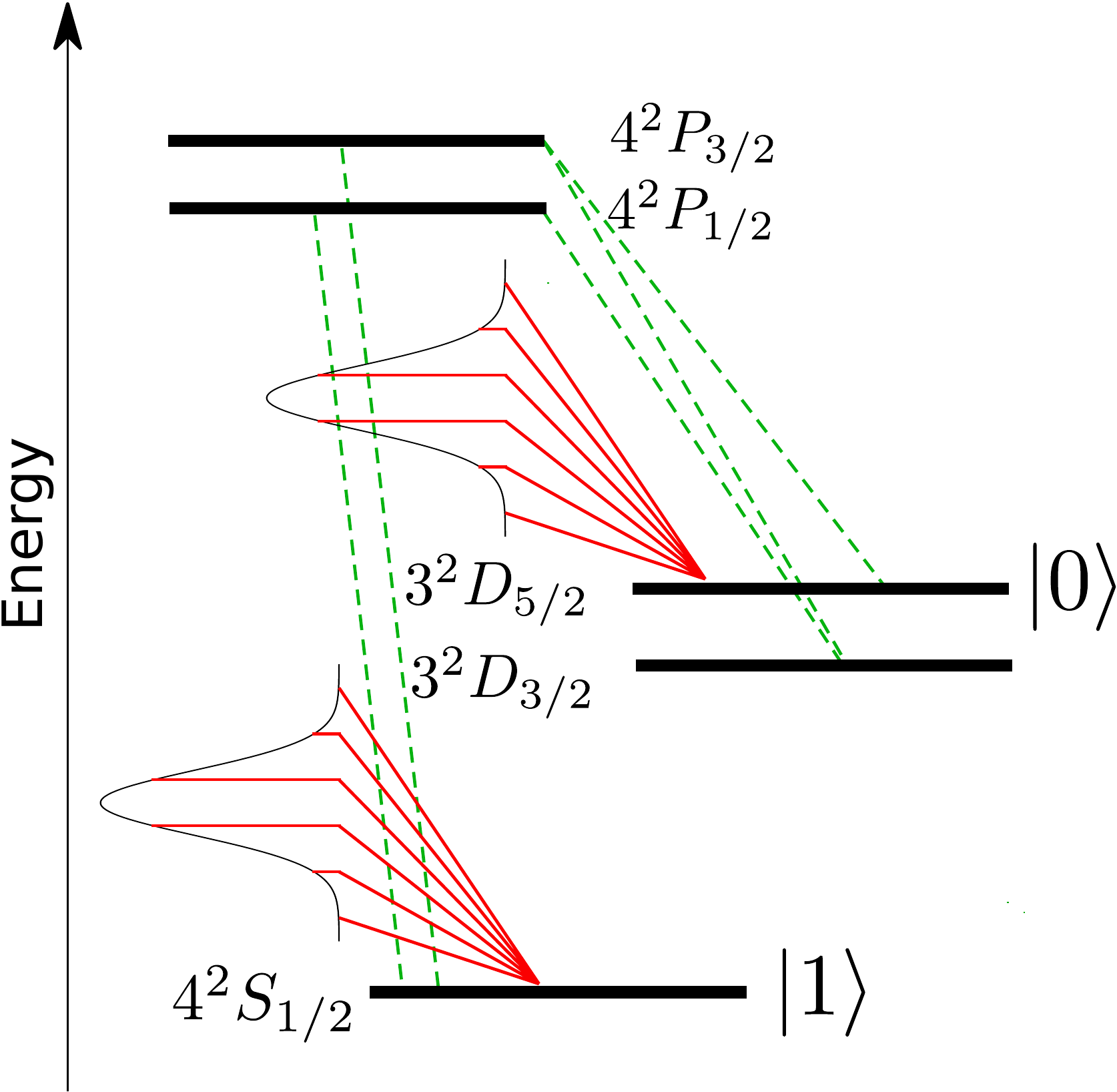}
  \end{minipage}
  \caption{
    The energy levels of the 
    $\Calcium$ 
    ion (black lines, not to scale) together with the frequency comb spectrum (red lines). The carrier frequency of the comb is far detuned from all 
  the transitions between the ion levels. The green dashed lines represent the dipole allowed transitions.
  }
  \label{fig:ion_levels}
\end{figure}

\section{Theoretical description of the gate implementation}
\label{sec:theory}
To find parameters for the suggested gate implementation,
we derive the evolution operator for trapped ions interacting with the EM field $\vec{E}(t, \hat{x})$ of the combs using the following Hamiltonian:
\begin{equation}
    \label{eq:full_hamiltonian}
    \begin{gathered}
        \mathcal{H} = \hat{H}_0 - \vec{E}(t, \hat{x}) \hat{\vec{d}} + 
        \sum \hbar \omega_\lambda \hat{a}_\lambda^\dagger \hat{a}_\lambda,\\
         \hat{H}_0 = \sum_\alpha \epsilon_\alpha|\alpha\rangle\langle\alpha|,\\
         \hat{\vec{d}} = \sum_{\alpha\beta} \vec{d}_{\alpha\beta}  |\alpha\rangle\langle\beta|,\\
        \hat{x} = x_0 + \sum_\lambda \frac{\eta_\lambda}{k_c} (\hat{a}_\lambda + \hat{a}^\dagger_\lambda).
    \end{gathered}
\end{equation}
Here $\hat{H}_0$ is the Hamiltonian of a single ion with the electronic levels $\alpha$, $\epsilon_\alpha$ are their corresponding energies, $\hat{\vec{d}} = (\hat{d}_x, \hat{d}_y, \hat{d}_z)$ 
is the ion dipole moment operator defined in the basis of the electronic states $\alpha,\beta$,
$\eta_\lambda$ is the Lamb-Dicke parameter of the normal mode $\lambda$, 
$k_c = \omega_c/c$ is the comb wavevector, and $\hat{a}_\lambda^\dagger,\hat{a}_\lambda$ are 
phonon creation/annihilation operators. 
For the energy level structure and the dipole-allowed transitions of the $\Calcium$ ion, see Fig.~\ref{fig:ion_levels} (see also \cite{NIST_ASD}).%} 

For our purposes, it is convenient to write the field 
of two combs $\vec{E}(t,\hat{x})$ acting on each ion
in the form where the contributions from each comb are grouped into pairs:
\begin{equation} \label{eq:two_pulses_field}
\vec{E}(t, x) = \sum_{k=1}^{N_\mathrm{pulses}} \mathcal{E}_k(t, x)\vec{u},
\end{equation}
\begin{multline}
\mathcal{E}_k(t, x) = E_{env}\left(t - kT-t_1 - \frac{x}{c}\right)e^{-i\omega_c(t - t_1 - \frac{x}{c})}\\ + 
E_{env}\left(t - kT - t_2 + \frac{x}{c}\right)e^{-i\omega_c(t - t_2 + \frac{x}{c})} + \mathrm{c. c.}.
\end{multline}
Here $T = \dfrac{1}{\nu_{rep}}$  is the time between the pulses, $t_{1, 2}$ are the delay times of the combs, $\vec{u}_{1, 2}$ are polarization vectors of the combs, and $\omega_c$ is the carrier offset frequency. The envelopes $E_{env}(t)$ of the comb pulses are smooth functions quickly decaying away from $t = 0$, in particular, we assume Gaussian shape $E_{env}(t) = E_\mathrm{peak} e^{-t^2/\tau^2}$. The terms $\mathcal{E}_k(t, x)$ contain contributions of the $k$-th 
pulses of both frequency combs. 
We can separate the contributions of different pairs of pulses because in our scheme the $k$-th
pulses from both combs come with the delay of tens of 
femtoseconds, whereas the interval between the adjacent pulses is tens of nanoseconds.

We use the following strategy to find the evolution operator of an ion in the field of the frequency combs. We find the evolution operator for each couple 
of pulses and then multiply all the operators for all couples. The free evolution between the couples is represented by the identity operator, so the evolution 
operator of the train of pulses reads
\begin{equation}
  \label{eq:u_as_product_aperiodic}
%  U_{total} = U_{N} U_\mathrm{free} U_{N-1} \dots U_\mathrm{free} U_{1},
  U_{total} = U_{N_\mathrm{pulses}} U_{N_\mathrm{pulses}-1} \dots U_{1},
\end{equation}
where each operator $U_k$ represents the action of the $k$-th couple of pulses with the field $\mathcal{E}_k$ in the interaction representation.

In the interaction picture, the unitary rotation $U_k$
can be found with the help of the perturbation theory based on Magnus expansion \cite{Magnus1954}. 
It is sufficient to consider the Magnus expansion up to the second order as the phase shift on the ion levels appears first in the second order.
Thus, the evolution operator in the interaction representation reads
\begin{equation}
  U_k = e^{X_k+Y_k},
\end{equation}
where $X_k$ and $Y_k$ are the first-order and the second-order contributions of the $k$-th couple of pulses. 

Using the Fourier image of the field $\mathcal{E}_k(\omega)$ of the couple of pulses 
and ignoring the contribution of the phonon modes,
one gets
\begin{equation}
  \label{eq:1_ord_magnus}
  (X_k)_{\alpha\beta} = -i\mathcal{E}_k(\epsilon_\alpha - \epsilon_\beta) (\vec{u}\vec{d})_{\alpha\beta}.
\end{equation}
\begin{widetext}
\begin{equation}
  \label{eq:2_ord_magnus}
   (Y_k)_{\alpha\beta} = 
   +i \int \frac{d\omega}{2\pi} 
   \mathcal{E}_k \left(\frac{\epsilon_\alpha - \epsilon_\beta}{2} - \omega\right)
   \mathcal{E}_k \left(\frac{\epsilon_\alpha - \epsilon_\beta}{2} + \omega\right)
   \sum_\gamma\frac{(\vec{u}\vec{d})_{\alpha\gamma}(\vec{u}\vec{d})_{\gamma\beta}}
   {\epsilon_\gamma - \frac{\epsilon_\alpha + \epsilon_\beta}{2} - \omega}.
\end{equation}
\end{widetext}
Eqs. \eqref{eq:1_ord_magnus} and \eqref{eq:2_ord_magnus} contain all possible contributions from the single-photon and two-photon processes except 
those involving the phonon modes
(for the latter, see Section~\ref{sec:errors}).
$(X_k)_{\alpha\beta}$ contains contributions from single-photon absorption and stimulated emission. In our setup, it is 
strongly off-resonant and can be neglected. 
$(Y_k)_{\alpha\beta}$ contains contributions from two-photon transitions, 
Raman transitions and photon forward scattering. Among these processes, 
two-photon absorption and emission can be neglected as they are off-resonant. Raman transitions are possible, and
they are one of the sources of gate infidelity. In particular, for each couple of pulses, there is a nonzero transition amplitude between the ion fine structure components, Zeeman sublevels or oscillatory levels. However, for a long train of pulses, these amplitudes interfere destructively with the appropriate choice of the repetition rate, train duration, and the intensity of combs. Thus, the only remaining effect is photon forward 
scattering which leads to the phase accumulated on each ion level. The phases can be directly found from the diagonal components 
of $Y_{\alpha \beta}$. 
For the ion at the position $x$, the acquired phase on the level $\alpha$ is 
\begin{widetext}
\begin{equation}
  \label{eq:phase_shift}
  \delta\theta_{\alpha}(x) = -iY_{\alpha\alpha} \approx  4\int \frac{d\omega}{2\pi} |E_{env}(\omega-\omega_c)|^2\left[1 + \cos{\left(\omega\left(t_1-t_2+\frac{2x}{c}\right)\right)}\right]
   \sum_\gamma\frac{(\epsilon_\gamma - \epsilon_\alpha)|(\vec{u}\vec{d})_{\alpha\gamma}|^2}
   {(\epsilon_\gamma - \epsilon_\alpha)^2 - \omega^2}.
\end{equation}
\end{widetext}
where $E_{env}(\omega)$ is the Fourier image of a single pulse envelope, and only the resonant contributions to the $Y_{\alpha\alpha}$ are kept.

\begin{figure}[h]
    \centering
    \includegraphics[width=\linewidth]{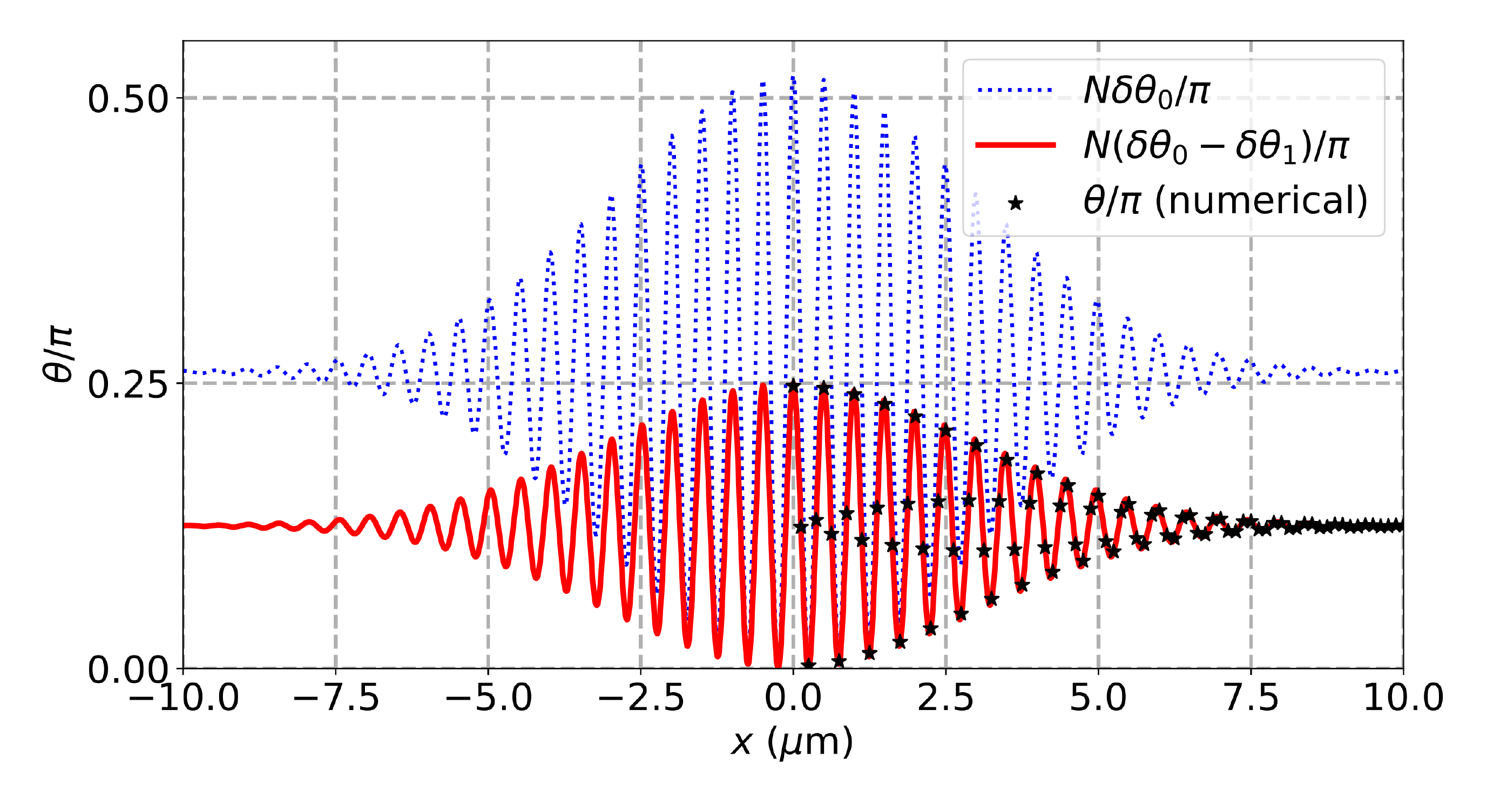}
    \includegraphics[width=\linewidth]{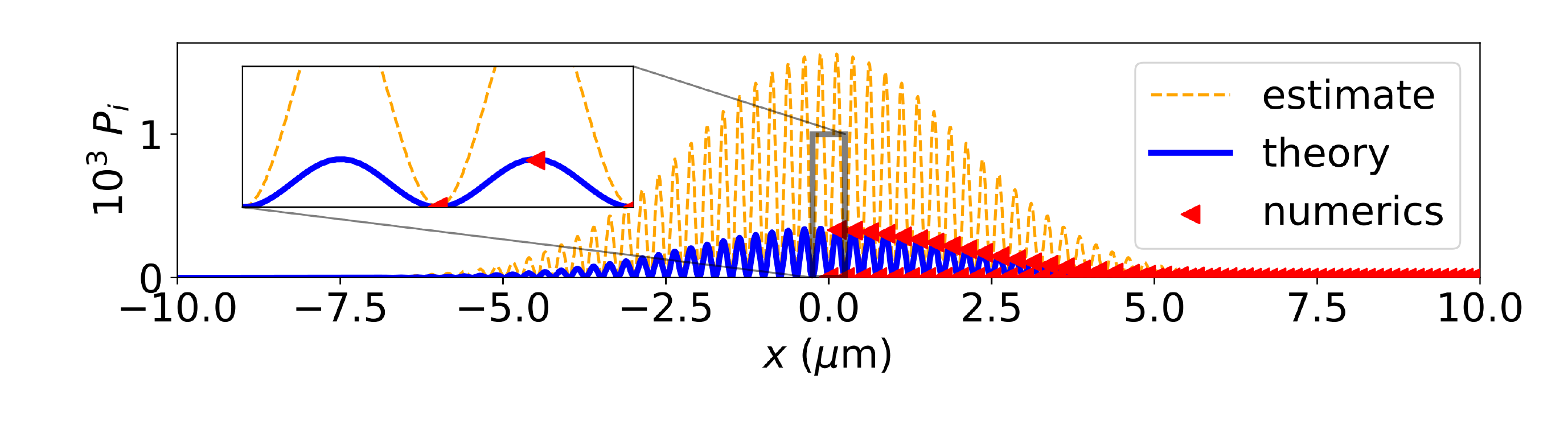}
    \caption{(a) The phases acquired on the ion qubit levels $|0\rangle = |D_{5/2}, m=-1/2\rangle$ and $|1\rangle = |S_{1/2}, m=-1/2\rangle$ after the action of 
    $N_\mathrm{pulses} = 200$ couples of combs pulses depending on the ion 
    coordinate $x$ The parameters of combs are taken from 
    Table~\ref{tab:combs_and_gate_params}. 
(b) %The upper estimate \eqref{eq:phonon_excitation_estimate}, the analytical 
    %expression \eqref{eq:phonon_excitation_exact} and the numerically 
    %simulated values for 
    The probability of the phonon mode excitation
    for the initial state 
  $|\psi_\mathrm{init}\rangle = \frac{1}{\sqrt{2}}\left(|0\rangle + |1\rangle\right)$. The orange dashed line represents the upper estimate 
  \eqref{eq:phonon_excitation_estimate}, the blue solid line represents 
  the analytical expression \eqref{eq:phonon_excitation_exact}, and 
  red triangles represent
  the numerically simulated values.
  }
    \label{fig:phases}
\end{figure}

The behavior of the phases $\delta\theta_{\alpha}(x)$ for the levels of the $\Calcium$ is shown in 
Fig.~\ref{fig:phases}. 
The phases $\delta\theta_{\alpha}(x)$ have 
oscillatory dependence on $x$ due to the interference of combs. They vanish away from the overlap region, and the phases approach the constant value. 
In the center of the overlap region (at $x = c(t_2-t_1)/2$), the phases are twice of the value far away from it. 
Thus, by applying $N_\mathrm{pulses}$ couples of pulses to the ion chain with the combs delays $t_1$ and $t_2$ such as $x_{tg} = c(t_2 - t_1)/2$,  one can implement the $R_z(2\theta)$ rotation on the target ion with the coordinate $x_{tg}$ and $R_z(\theta)$ rotation on all other ions, where
\begin{equation}
  \label{eq:total_phase}
  \theta = N_\mathrm{pulses}\delta\theta(x_{tg}) = N_\mathrm{pulses}(\delta\theta_{1}(x_{tg}) - \delta\theta_0(x_{tg})/2, 
\end{equation}
with the indices $0, 1$ denoting the qubit levels and the gate duration of $N_\mathrm{pulses}T$. 

With Eq.~\eqref{eq:phase_shift}, we calculate the phase shifts for the $4S_{1/2},\,m=-1/2$ and $3D_{5/2},\,m=-1/2$ levels of the 
$\Calcium$ ion. The energy levels (see Fig.~\ref{fig:ion_levels}) 
are taken from the NIST database \cite{NIST_ASD}, and we calculate
the transition dipole moments from the decay rates taken from the NIST database. 
The parameters of combs are taken from 
Table~\ref{tab:combs_and_gate_params}. 
The field polarization is taken as $\vec{u} = (0, 0, 1)^T$ (parallel to the quantization axis) for simplicity, and
the value of the electric field is chosen in order to achieve the qubit rotation speed of $\frac{d\theta}{dt} = \pi/16 \,\mu\text{s}^{-1}$. This
allows performing a $\pi/2$ rotation per $8\,\mu s$ with $800$ combs pulses, which is comparable with the duration of single-qubit gate
implemented with continuous-wave lasers.

Also, by the appropriate choice of the comb wavelength and intensity, our method can be adjusted for the 
$\Strontium$, $\Barium$, and $\Radium$ ions which have a similar structure of the energy levels 
(see Fig.~\ref{fig:ion_levels}).

\begin{table}
    \centering
    \begin{tabular}{  |c|c|c|  } 
    \hline
%     & Designation & Value  \\
    \multicolumn{3}{|c|}{\textbf{Combs parameters}}  \\
    \hline
    carrier wavelength & $\lambda_c$ & $1000 $ nm  \\
    pulse duration & $\tau$ & $20$ fs \\ 
    repetition rate & $\nu_{rep}$ & $100$ MHz  \\ 
    polarization& $\vec{u} $ & $(0, \, 0, \, 1)^T$ \\ 
    peak electric field & $ea_0E_\mathrm{peak}/\hbar$ & $\approx 4.405 $ THz  \\ 
    \hline
    \multicolumn{3}{|c|}{\textbf{Trap parameters}}  \\
    \hline
    Axial frequency    & $\omega_{ax}/(2\pi) $ & 600\,\text{kHz}\\
    Lamb-Dicke parameter    & $\eta$   & 0.09\\
    \hline
    \multicolumn{3}{|c|}{\textbf{Simulated gate parameters}}  \\
    \hline
    angle per couple of pulses & $\delta\theta(x=0)$ & $\pi/1600$  \\
    qubit rotation speed & $\frac{d\theta}{dt}$ & $\pi/16 \,\mu\text{s}^{-1}$\\
%    gate time & $t_{gate}$ &  $8 \mu s$  \\
%    number of pulses & $N_\mathrm{pulses}$ &  $800$  \\
    infidelity & $1-F$ & $4\cdot 10^{-4}$ \\ 
    \hline
    
%    beam waist & $w$ & $30$ $\mu$m  \\ 
%    \hline
%    comb power & $W$ & $0.3$ W  \\ 
%    \hline
    
    \end{tabular}
    \caption{Two optical frequency combs parameters for single-qubit trapped ions gates.}
    \label{tab:combs_and_gate_params}
\end{table}

\section{Gate errors}
\label{sec:errors}

Assuming the ideal combs field, the intrinsic sources of the gate error are as follows:
\begin{enumerate}
  \item Unwanted rotations of the non-target qubits,
  \item Photon scattering,
  \item Transitions between the electronic or vibrational levels.
\end{enumerate}

The unwanted rotations of the non-target qubits (the crosstalk) occur because the phase shifts of the non-target qubits are not exactly equal to each other. 
This leads to the incomplete cancellation of their phases when 
the global rotations are applied (see Fig.~\ref{fig:setup}(b, c)). The dominant contribution to the error comes from the 
nearest neighbors of the target qubit. %, and it can be calculated with the help of the Eq.~\eqref{eq:phase_shift}. 
Given that the target qubit rotates by $2\theta$, the neighbor ones rotate by $\theta + \delta\theta$. For the 
other ions, one can neglect the incomplete phase cancellation due to the rapid decay of the phases oscillations in Eq.~\eqref{eq:phase_shift}.
%safely assume that they rotate exactly by $\theta$. 
Then, the contribution to the gate infidelity can be estimated as $\delta\theta^2$. For the inter-ion distance of $10 \mu\mathrm{m}$, 
we calculate the phase difference $\delta\theta$ with the help of Eq.~\eqref{eq:phase_shift} and find that $\delta\theta \sim 10^{-2}$ 
for $\theta \sim 1$, 
which leads to the contribution to the gate infidelity of $\sim 10^{-4}$.

For photon scattering, we estimate its contribution into the gate infidelity as the 
mean number of scattered photons (both elastically and inelastically) by all ions after the action of the train of pulses.
Given that the full scattering cross section of a photon with the frequency $\omega$ is $\sigma(\omega)$, %the total number of 
we calculate the mean number of photons scattered by a single ion from a single comb pulse $E(t)$ as
\begin{equation}
  \label{eq:n_scattered}
  N_\mathrm{scattered} = \frac{\epsilon_0c}{\hbar}\int_0^\infty\frac{d\omega}{\pi} \frac{\sigma(\omega)}{\omega} |E(\omega)|^2.
\end{equation}
The cross section $\sigma(\omega)$ is given by the Kramers-Heisenberg formula \cite{Berestetskii2012} integrated over the scattered photon angles and summed over all possible 
final states. In the Kramers-Heisenberg formula, we carefully take into account the finite lifetime of the intermediate $P_{1/2}$ and $P_{3/2}$ levels
in order to avoid singularities in the integral. The resulting probability to emit a photon per pulse at the considered parameters 
does not exceed $10^{-9}$, which leads to the contribution to infidelity of $\sim 10^{-6}$ per ion for $\sim 1000$ pulses.

Then, we analyze the contribution of the transitions between the electronic and the vibrational levels. We claim that the most important processes
contributing to the gate error are the following Raman transitions:
\begin{enumerate}[label=(\alph*)]
  \item Between different fine structure sublevels: $D_{5/2} \to D_{3/2}$,
  \item Between different Zeeman components of the qubit manifolds,
  \item Between different oscillatory levels (equivalent to the phonon creation and annihilation).
\end{enumerate}
Among the other possible transitions, the excitation of the 
short-living $P_{1/2}$ and $P_{3/2}$ levels contributes to the photon scattering discussed above, as 
the excited $P_{1/2}$ and $P_{3/2}$ levels quickly decay with the emission of a photon.
%the first-order excitation of the $P_{1/2}$ and $P_{3/2}$ levels 
Also, the quadrupole and magneto-dipole transitions between the $S$ and $D$ levels and the two-photon 
absorption and emission processes are strongly off-resonant and thus can be neglected.

In contrast, for the processes (a)-(c), the single-pulse transition 
amplitudes have comparable magnitude with the phase shifts defined by \eqref{eq:phase_shift} 
because of the large spectral width of the pulses exceeding even the fine splitting. However, the total contribution for the train of pulses obtained 
by the summation of all pulse amplitudes remains small due to the destructive interference between the pulses and does not 
accumulate with the increasing $N_\mathrm{pulses}$. 

The detailed calculations for the processes (a) and (b) are presented in the Appendix A. 
There, we found the transition amplitudes for a train of pulses by direct summation of all the 
single-pulse amplitudes. 

For the transition amplitude $D_{5/2}\to D_{3/2}$, the total transition probability remains as small as 
$\sim 1/N_\mathrm{pulses}^2 \sim 10^{-6}$ providing 
that the transition frequencies between the qubit $D_{5/2}$ level  
and $|D_{3/2},m\rangle$ levels are not the integer multiples of the repetition rate. 
For the transitions between the Zeeman sublevels of the $D_{5/2}$, we estimate the transition probability as 
$\left(\frac{\nu_{rep}}{2\pi N_\mathrm{pulses}\nu_{z}}\right)^2 \sim 10^{-4}$, where $\nu_z$ is the Zeeman splitting.
%The detailed calculation is presented in the Appendix~\ref{appendix:non_qubit_transitions}.%, and the results are as follows.

For the process (c), %the 
we calculate the probability of the phonon mode excitation with the help of the effective Hamiltonian for the qubit levels and phonon modes 
(see details in the Appendix~\ref{appendix:phonon_excitation}). 
We derive the effective Hamiltonian using the fact that the normal mode oscillation period is much larger than 
the single pulse duration and considerably larger than the period between 
pulses. 
Given that the ion is initially in the state
$|\psi_{init}\rangle = c_0|0\rangle + c_1|1\rangle$,
the estimate for the phonon excitation probability reads
(see Eq.~\eqref{eq:phonon_excitation_exact})
\begin{equation}
  \begin{gathered}
    \label{eq:phonon_excitation_estimate}
    P_i = |c_0|^2 P_{i0} + |c_1|^2 P_{i1} < P_{i0},\\
    P_{i\alpha} < \frac{c^2}{\omega_c^2}\left(\frac{\p\delta\theta_\alpha}{\p x}\right)^2\sum_{s} \frac{4|\eta_{is}|^2}{\omega_{s}^2T^2},
  \end{gathered}
\end{equation}
where $i$ is the ion number, %$\alpha$ is the ion electronic state, 
$s$ enumerates the phonon modes, 
and $\eta_{is}$ are the Lamb-Dicke parameters. The first inequality holds because $P_{i0} > P_{i1}$.

We estimate the phonon excitation probability for a single ion in a trap with the axial frequency $\omega_{ax} = 600\,\mathrm{kHz}$ (with the Lamb-Dicke parameter for the axial mode $\eta = k_c\sqrt{\hbar/(2m\omega_{ax})} = 0.09$). We assume that the center of the interference peak deviates by no more than $30\,\mathrm{nm}$ from the ion position $x$. Using these conditions, the parameters from Table \ref{tab:combs_and_gate_params}, 
and the inequality \eqref{eq:phonon_excitation_estimate}, we find that the phonon excitation probability does not exceed $3\cdot 10^{-4}$.

%For the considered gate parameters and for the case of a single ion in a trap with a realistic value of the axial frequency 
%$\omega_{ax} = 600\mathrm{kHz}$ (with the Lamb-Dicke parameter for the 
%axial mode $\eta = k_c\sqrt{\hbar/(2m\omega_{ax})} = 0.09$),
%the phonon excitation probability does not exceed $3\cdot 10^{-4}$ providing that the ion coordinate $x$ deviates by no more than $30\mathrm{nm}$ from the 
%center of the interference peak.

All the discussed errors are presented in Table~\ref{tab:error_sources}. The dominant contributions are the crosstalk
and the phonon excitation error. 
We estimate the total contribution for the suggested parameters as $4\cdot 10^{-4}$.

%\begin{TAB}(r,1cm,2cm)[5pt]{|c|c|}{|c|c|c|}% (rows,min,max)[tabcolsep]{columns}{rows}
%hi & tall one    \\
%hi & medium one  \\
%hi & standard one\\
%\end{TAB}
%\begin{widetext}
\begin{center}
  \begin{table}
      \begin{tabular}{ | c | c | } 
      \hline
      \textbf{Error source}  & \textbf{Contribution}  \\
      \hline
      Crosstalk & $10^{-4}$  \\
      \hline
      Photon scattering & $10^{-5}$ \\ 
      \hline
      Transitions to non-qubit Zeeman sublevels &  $10^{-6}$ \\ 
      \hline
      Transitions to $D_{3/2}$ sublevels &  $10^{-6}$ \\
      \hline
      Phonon excitation  & $3\cdot 10^{-4}$ \\ 
      \hline
      \textbf{Total error} & $4\cdot 10^{-4}$\\
      \hline
      \end{tabular}
      \caption{The contributions to the frequency-comb 
      based gate error discussed in Section~\ref{sec:errors}.}
      \label{tab:error_sources}
  \end{table}
\end{center}
%\end{widetext}

%The analysis of the Appendix~\ref{appendix:errors} shows that the dominant sources of error is the excitation of the vibrational modes. 

%For $\Calcium$ ion, the subsequent analysis of errors shows that the high gate fidelity of the $\pi/2$ rotation can be obtained for the $8-10\mu s$ gate 
%duration, which corresponds to 800-1000 combs pulses. 

\section{Numerical modelling}
We verify the theoretical considerations of Sections \ref{sec:theory} and \ref{sec:errors}
with the numerical simulation 
of the $\Calcium$ ion in the field of two counter-propagating frequency combs.
In the simulation, we consider the low-lying $S$, $P$, and $D$ levels of the $\Calcium$ ion, 
the interaction of the internal degrees of freedom of the ion with one vibrational mode of the ion chain, and
radiative decay of the short-living $P$ levels. 
%and the quadrupole transitions between $S$ and $D$ levels.

For the verification, we simulate the quantum dynamics of the ion using the master equation in the Lindblad form \cite{Gorini76a,Lindblad76a}:
\begin{equation}
  \label{mast_eq} \frac{\hat{d} \hat{\rho}}{d t} = -\frac{i}{\hbar}[\hat{\mathcal{H}}, \hat{\rho}] + \mathcal{L}\hat{\rho}, 
\end{equation}
where $\hat\rho$ is the density matrix of the ion-phonon system, 
$\hat{\mathcal{H}}$ is the Hamiltonian \eqref{eq:full_hamiltonian}, 
and $\mathcal{L}$ is a spontaneous relaxation superoperator \cite{FicekBook,AgarwalBook}: 
%This superoperator reads (see~\cite{FicekBook,AgarwalBook})
\begin{equation}
  \label{Lind_op} 
  \mathcal{L}\hat{\rho} = -\sum_{c} \frac{\gamma_{c}}{2}\left ( \hat{\rho} \hat{S}_c^+\hat{S}_c^- 
  + \hat{S}_c^+\hat{S}_c^- \hat{\rho} - 2\hat{S}_c^-\hat{\rho} \hat{S}_c^+ \right ), 
\end{equation}
Here $c$ enumerates the set of all possible spontaneous decay channels 
for the considered energy levels, $\gamma_c$ are the spontaneous decay rates,
and $\hat{S}_c^\pm$ are the excitation/deexcitation operators. 
The actual simulations are performed in the
interaction representation. We use the QuantumOptics.jl package \cite{Kramer2018} written in the Julia language \cite{Bezanson2017julia}. 

%\textcolor{red}{
%We include the following levels into the simulation: $|4S_{1/2}, m=\pm 1/2 \rangle,$ and $|4P_{1/2}, m=- 1/2\rangle,$ $|4P_{3/2}, m=- 1/2\rangle,$ $|3D_{3/2}, m=- 1/2\rangle,$ $|3D_{5/2}, m=- 1/2\rangle.$ The $\sigma_z$-polarization of combs is considered, and two quadrupole and all the dipole allowed terms between levels are included into electronic part of the full Hamiltonian. The data for decay rates and other transition parameters are taken from NIST database \cite{NIST_ASD} and from the works \cite{Chwalla2009}-\cite{Huang2012}.}
%

To model the gate described above, 
%we solve numerically the quantum master
%equation \eqref{mast_eq}. 
we consider the ion at $t=0$ described by the normalized phonon-ion wavefunction
\begin{equation}
  |\psi_{init} \rangle = \left (c_0 \,  |0\rangle_q + c_1 |1\rangle_q \right )\otimes|0\rangle_{vib},
\end{equation}
where $|n \rangle_{vib}$ is the $n^{th}$ Fock state of the vibrational mode, $|0\rangle_q$ and $|1\rangle_q$ are the qubit states encoded in 
the $3D_{5/2},\,m=-1/2$ and $4S_{1/2},\,m=-1/2$ levels respectively,
and $c_0 = c_1 = 1/\sqrt{2}$. The initial density matrix is 
$\hat{\rho}_{init} = %\hat{\rho}(t=0) 
|\psi_{init}\rangle\langle\psi_{init}|$. Then, we
calculate the evolution under $200$ pairs of combs pulses with the model parameters of Table~\ref{tab:combs_and_gate_params} for different ion equilibrium positions
$x_0$.
 
For the final density matrix $\hat{\rho}_\mathrm{fin}$ in the interaction 
representation, we calculate the phase difference $\theta$ acquired between 
the qubit levels with the expression
%
%The induced phase $\theta_N$ after $N$ pulses is found as the argument of the complex non-diagonal $4S_{1/2}-3D_{5/2}$ density matrix element with the following formula:
\begin{equation}
  \theta = \Arg \left ( \langle \psi_0 |\hat{\rho}_\mathrm{fin} | \psi_1 \rangle \right ),
\end{equation}
where $|\psi_\alpha\rangle = |\alpha\rangle_q \otimes |0 \rangle_{vib}$. 
Also, we calculate the populations $\rho_{\gamma\gamma}$ 
of the non-qubit states. 

The numerical results match with the theoretical predictions of
Sections \ref{sec:theory} and \ref{sec:errors}. The acquired qubit phase 
coincides with Eqs.~\eqref{eq:phase_shift} and \eqref{eq:total_phase} with
the discrepancy of $\sim 10^{-3}$ (see Fig.~\ref{fig:phases}(a)), 
which we associate with the higher order terms of the Magnus expansion.
This discrepancy does not contribute to the gate error, as it indicates only the 
inaccuracy of the analytical expression for the phase~\eqref{eq:phase_shift}
and can be avoided with the appropriate calibration of the combs power.
The probability of the phonon excitation also matches with the prediction
of Eq.~\ref{eq:phonon_excitation_exact} (see Fig.~\ref{fig:phases}(b)). 
The excitation of the non-qubit electronic levels remains on the level of $10^{-6}$ in 
agreement with the estimates of Section \ref{sec:errors}.

\section{Conclusion}
We suggest a new method for single-qubit addressing in the trapped ion 
linear chains based on the quadratic Stark shift of two counter-propagating
frequency combs. The implementation requires only global laser beams and is applicable for the ion chain of any length. To realize a single-qubit 
gate on a target qubit, one should adjust the time delay between the
combs pulses so that the target ion lies in the overlap region between the pulses and apply the train of combs pulses 
and additional continuous-wave laser fields. 
We present the detailed calculation of the gate parameters for $\Calcium$ ions and 
find that the single-qubit rotations can be implemented with the low infidelity ($\sim 4\cdot10^{-4}$) at moderate gate times 
($\sim 8 \mu s$). 
Also, our method can be extended for the isoelectronic 
$\Strontium$, $\Barium$ and $\Radium$ ions by the appropriate choice of the setup parameters.
We believe that our findings provide useful insight into quantum operations in trapped ion chains.

%\section*{Acknowledgments}

%This work was funded in part by the EPSRC under Grant No. EP/R006474/1. We gratefully acknowledge helpful discussions with P. Crivelli, S. D. Hogan and R. Pohl. We would like to thank MPCDF Garching for computation time.

\nolinenumbers
%\clearpage
\bibliography{lib.bib}

\appendix
\section{Leakage to the non-qubit ion levels}	

  \label{appendix:non_qubit_transitions}

  First, we analyze the leakage to the non-qubit ion levels. They can belong to either $S_{1/2}$, $D_{3/2}$ or $D_{5/2}$ manifolds. For the qubit 
state $0$ belonging to the $D_{5/2}$ manifold, leakage is possible to the
$D$ sublevels, and for the qubit state $1$ belonging to
the $S_{1/2}$ manifold, the leakage is possible to the non-qubit $S_{1/2}$ component.

Assuming that the ion is in the qubit state $\alpha$, we find the transition amplitude due to the action of the train of pulses 
to the non-qubit state $\gamma$. It can be approximated as the sum of transition amplitudes associated with the action of each couple of pulses. 
Each of these processes is a two-photon process corresponding to the absorption and emission of a photon, therefore, the amplitude by
the $k$-th pulse can be written as $a_k = a_0e^{i(\epsilon_{\gamma} - \epsilon_{\alpha})kT}$. Therefore, the total amplitude reads
\begin{equation}
  \label{eq:leakage_amplitude}
  |a_{tot}| = \left|\sum_{k=0}^{N_\mathrm{pulses}-1} a_0e^{i(\epsilon_{\gamma} - \epsilon_{\alpha})kT}\right| <
  \frac{|a_0|}{|\sin{[(\epsilon_\alpha-\epsilon_\gamma)T/2]}|}.
\end{equation}
The dominant contribution to $a_0$ comes from the second order of the Magnus expansion. 
All these amplitudes 
%per a single pulse 
have the same order
as the phase shifts per pulse. For the gate times and repetition rates that we consider, 
$a_0 \sim \theta_\mathrm{gate}/N_\mathrm{pulses} \sim 10^{-3}$. 

%For the case of the leakage from the qubit state 
When the leakage occurs from the qubit state $0$
to the non-qubit fine-structure $D_{3/2}$ sublevel,
%$0$ ($D_{5/2}$) to $D_{3/2}$, 
the transition frequency is 
$\epsilon_{D_{5/2}} - \epsilon_{D_{3/2}} = 2\pi\cdot 1.8287\,\mathrm{THz}$. 
Keeping the repetition rate $\nu_{rep}$ unequal to an integer fraction of 
$(\epsilon_{D_{5/2}} - \epsilon_{D_{3/2}})/(2\pi)$, 
%is not an integer multiple of $2\pi\nu_{rep}$, 
one can easily ensure that
$\sin{[(\epsilon_{D_{5/2}} - \epsilon_{D_{3/2}})T/2]}| \sim 1$. For example, for the repetition rate of $100\,\mathrm{MHz}$, the energy difference can be 
represented as
%which is considered in the main text, 
$\epsilon_{D_{5/2}} - \epsilon_{D_{3/2}} = 2\pi(k + \Delta k)\nu_{rep}$, where $k = 18{,}287$ and $\Delta k = 0.34$. 
At these conditions, the probability of the transition to the $D_{3/2}$ sublevels remains as small as $\sim \theta_{gate}^2/N_{\mathrm{pulses}}^2 \sim 10^{-6}$.

For the case of the leakage from the qubit state to other Zeeman sublevels of the same manifold, the transition frequency equals the Zeeman splitting. 
Assuming the Zeeman splitting of $\nu_{z} = 2\,\mathrm{MHz}$ and using Eq.~\eqref{eq:leakage_amplitude}, one gets the transition probability below 
$\left(\frac{\nu_{rep}}{2\pi N_\mathrm{pulses}\nu_{z}}\right)^2 \sim 10^{-4}$.

\section{Excitation of the vibrational levels}
\label{appendix:phonon_excitation}
The transitions between the vibrational levels can be taken into account in the following way. 
As femtosecond pulses are much faster than the periods of the ion motion, 
one can neglect the time dependence of the operator $\hat{x}$ for each pulse. This allows expressing the evolution operator of the $k$-th pulse through the 
$x$-dependent evolution operator $U_k(x)$:
\begin{equation}
  U_k = U_k(\hat{x}(t_k))
\end{equation}
As the evolution operator for each pulse is approximately diagonal, 
for two qubit levels it takes the form
\begin{equation}
  U_k = 
  \begin{pmatrix}
    e^{i\delta\theta_0(\hat{x}(t_k))} & 0\\
    0 & e^{i\delta\theta_1(\hat{x}(t_k))}\\
  \end{pmatrix}
\end{equation}
The phases per pulse are small, therefore, we approximate the evolution of the wavefunction by a contiuous function. Thus, the evolution can be described by 
the effective Hamiltonian
\begin{equation}
H_\mathrm{eff} = \frac{1}{T}
  \begin{pmatrix}
    \delta\theta_0(\hat{x}) & 0 \\
    0 & \delta\theta_1(\hat{x})
  \end{pmatrix}.
\end{equation}
In the Lamb-Dicke regime, 
the expansion of the phases in deviations from the equilibrium position reads
%one can expand the phases as
\begin{equation}
  \delta\theta_\alpha(\hat{x}_i) = \delta\theta_\alpha(x_0) + \frac{1}{k_c}\frac{\p\delta\theta_\alpha}{\p x}\sum_s \eta_{is}(a_{s} + a^\dagger_{s}).
\end{equation}
Finally, the probability of phonon excitation (assuming that the gate acts on the $i$-th ion being in the level $\alpha$)
can be easily calculated with the first-order perturbation theory: 
%It results in 
\begin{multline}
  \label{eq:phonon_excitation_exact}
  P_{i\alpha} =\frac{1}{k_c^2}\left(\frac{\p\delta\theta_\alpha}{\p x}\right)^2\sum_{s} \frac{ |\eta_{is}|^2|e^{i\omega_{s} t_g} - 1|^2}{\omega_{s}^2T^2}\\ < \frac{1}{k_c^2}\left(\frac{\p\delta\theta_\alpha}{\p x}\right)^2\sum_{s} \frac{4|\eta_{is}|^2}{\omega_{s}^2T^2}.
\end{multline}
One can see that the probability is proportional to $\left(\frac{\p\delta\theta_\alpha}{\p x}\right)^2$. Therefore, for the ion at the center of the pulses overlap
region, it vanishes. Also, for the considered gate parameters and for the case of a single ion in a trap with a realistic value of the axial frequency 
$\omega_{ax} = 600\,\mathrm{kHz}$ (with the Lamb-Dicke parameter $\eta = k_c\sqrt{\hbar/(2m\omega_{ax})} = 0.09$),
the phonon excitation probability does not exceed $3\cdot 10^{-4}$ 
provided that the ion coordinate $x$ deviates no more than 
by $30\,\mathrm{nm}$ from the 
center of the overlap region.

\end{document}